# Architectures of Soft Robotic Locomotion Enabled by Simple Mechanical Principles

Liangliang Zhu [a, c], Yunteng Cao [a], Yilun Liu [b], Zhe Yang [a], Xi Chen [c*]

[a] International Center for Applied Mechanics, State Key Laboratory for Strength and Vibration of Mechanical Structures, School of Aerospace, Xi'an Jiaotong University, Xi'an 710049, China

[b] State Key Laboratory for Strength and Vibration of Mechanical Structures, School of Aerospace, Xi'an Jiaotong University, Xi'an 710049, China

[c] Columbia Nanomechanics Research Center, Department of Earth and Environmental Engineering, Columbia University, New York, NY 10027, USA

**Abstract**

In nature, a variety of limbless locomotion patterns flourish from the small or basic life form (*Escherichia coli*, the amoeba, etc.) to the large or intelligent creatures (e.g., slugs, starfishes, earthworms, octopuses, jellyfishes, and snakes). Many bioinspired soft robots based on locomotion have been developed in the past decades. In this work, based on the kinematics and dynamics of two representative locomotion modes (i.e., worm-like crawling and snake-like slithering), we propose a broad set of innovative designs for soft mobile robots through simple mechanical principles. Inspired by and go beyond existing biological systems, these designs include 1-D (dimensional), 2-D, and 3-D robotic locomotion patterns enabled by simple actuation of continuous beams. We report herein over 20 locomotion modes achieving various locomotion functions, including crawling, rising, running, creeping, squirming, slithering, swimming, jumping, turning, turning over, helix rolling, wheeling, etc. Some of them are able to reach high speed, high efficiency, and overcome obstacles. All these locomotion strategies and functions can be integrated into a simple beam model. The proposed simple and robust models are adaptive for severe and complex environments. These elegant designs for diverse robotic locomotion patterns are expected to underpin future deployments of soft robots and to inspire series of advanced designs.

---

[*] Corresponding author at: Columbia Nanomechanics Research Center, Department of Earth and Environmental Engineering, Columbia University, New York, NY10027, USA. Tel.: +1 2128543787.
E-mail address: xichen@columbia.edu (X. Chen).

## 1. Introduction

Conventional locomotion systems can be very unstable in complex or varying environments [1], have only limited motions, and require complex manufacturing processes involving a large number of rigid components and a larger number of discrete joints. For enhanced adaptability toward complex terrains, biomimetic hard robots have been developed in the past decades, which were composed of a wide set of hard components, joints, sensors, sophisticate yet complex control algorithms. The recent CHEETAH robot [2] or the Atlas robot [3] may stand as some of the best examples of hard robots to date.

Contrary to artificial locomotion systems, the biological organisms' locomotion mechanisms are usually flexible, agile, and highly adaptive to rough and varying environments and terrains; they are ubiquitous in nature from single cell (e.g. bacteria, the amoeba, sperm cells, etc.) to large and intelligent animals (e.g. snakes, fishes, octopuses, etc.). Inspired by the natural (limbless) locomotion (or motion), the development of soft robots has become a focal point in recent years [4-12], which aimed to achieve various soft movements and motions that are normally inaccessible to conventional "hard" robots. These studies, in turn, also stimulated the fundamental understandings of a few locomotion mechanisms or patterns of the biological organisms.

Highly potential applications for soft-robotic systems may range from medical endoscopy [13], industrial inspection, search-and-rescue operations, wearable devices [11, 14, 15], to unknown terrain exploration. Some exciting soft robot prototypes have been constructed in the past decades. For example, a quadrupedal soft robot with five actuators and a pneumatic valving system was constructed with crawling and undulation gaits [5]. Another example was a snake-like soft robot made of bidirectional fluidic elastomer actuators with passive wheels to generate the necessary frictional anisotropy [9]. Inspired by the ballistic movements of caterpillar musculature, a rolling robot [16] was developed and was able to achieve relatively high speed and efficiency. Marchese et al. reported recently a soft robotic fish actuated by soft fluidic actuators [12] with very flexible and rapid motions (capable of escape maneuvers). And Wehner et al. [17] has demonstrated lately the first autonomous, untethered, entirely soft robot powered by monopropellant decomposition. More intriguingly, at small scales, the rapid developing micro-fabrication techniques and the growing understandings of locomotion mechanisms of some small biological systems have advanced the development of artificial, bioinspired micro-soft robots [18, 19].

Various natural locomotion modes have inspired their counterpart biomimetic robots, such as the caterpillar or worm inspired locomotion [10, 16, 20-23], multi-gait quadruped locomotion [5, 24], snake inspired locomotion [8, 9, 25], and jellyfish or fish inspired swimming [4, 12]. With racing developments of sensors, actuators and 3-D fabrications [5, 9, 12, 14, 24, 26-29] (see a review in Ref. [15]), the reported architectures of soft locomotion may have possessed only a corner in the nature's limbless locomotion kingdom. In particular, the existing locomotion mechanisms and designs are still quite complex, and contrary to the fact that certain Nature's machineries may be designed simple.

In this work, based on two simple representative locomotion modes (i.e., worm-like crawling [30] and snake-like slithering [31] of a beam model), we propose a broad set of new designs for soft robotic locomotion through simple beam models, and demonstrate them through finite element simulations. Over 20 locomotion modes are reported herein, achieving various locomotion functions, including crawling, creeping, squirming, slithering, swimming, running, jumping, turning, turning over, helix rolling, wheeling, etc. Strategies for implementing some of these modes in practice are discussed. Inspired by and go beyond existing biological mobile systems, these designs may help to further understand and exploit the nature's various offers of limbless locomotion strategies.

The organization of the paper is as follows. In Section 2, two fundamental locomotion principles, the worm-like crawling [30] and snake-like slithering [31] proposed in our previous works, are briefly discussed. Based on these two basic locomotion patterns, in Section 3, a series of new deigns of soft robotic locomotion strategies are established and verified using FEM (finite element method) simulations. We compare our designs with some other existing soft locomotion in Section 4, concerning the flexibility, adaptability, and efficiency. The practical implementation methods are presented in Section 5, followed by concluding remarks.

## 2. Two representative soft robotic locomotion patterns

In this section, we review two of the most fundamental beam-based limbless locomotion patterns, i.e., the worm-like crawling [30] and snake-like slithering. Their kinematic mechanisms are briefly introduced.

### 2.1 Out-of-plane worm-like crawling locomotion

Shown in Fig. 1a is a beam model (length $L$, height $h$) for the worm-like crawling locomotion. The two ends of the beam are rounded with a radius of $h/2$ to allow stable contact with the ground. The friction coefficient is $\mu_1$ between the left end of the beam and the ground, and $\mu_2$ between the rest ($\mu_1<\mu_2$) of the beam and the ground. Through periodic bending and unbending of the beam (Fig. 1b), the differential friction force generated at the two ends can lead to a net crawling locomotion (Fig. 1c) [30]. Note that asymmetric friction force can be introduced in many ways, and here we use nonuniform friction coefficient as an illustration. Alternatively, the friction property can be uniform and asymmetry can be introduced through variation in cross-section, density, and/or different loading history on left/right sections, etc.

In practice, bending (or unbending) of the beam can be actuated through shape memory polymers [27], dielectric elastomers [26, 32], electroactive polymers [33], bimorph actuator [34-37], or pneumatic/fluid pressurization [5, 11, 38, 39], etc., discussed in Section 5. In simulation, repetitive beam bending and unbending can be achieved, by applying an extension strain, $T_1 = T_{max} \cdot 2\bar{t}$ when $0 \leq \bar{t} < 0.5$ and $T_{max} \cdot 2(1-\bar{t})$ when $0.5 \leq \bar{t} < 1$ ( $\bar{t} = t/P$ , where $P$ is the period), to the upper surface, and a contraction strain ($T_2 = -T_1$) to the lower surface of the beam. Without losing generality, the deformed configuration of the beam (inset of Fig. 1 b) can be regarded as a circular arc with central angle $\theta(\bar{t}) = 2LT_1/h$, and radius $R = L/\theta(\bar{t})$ [30, 40].

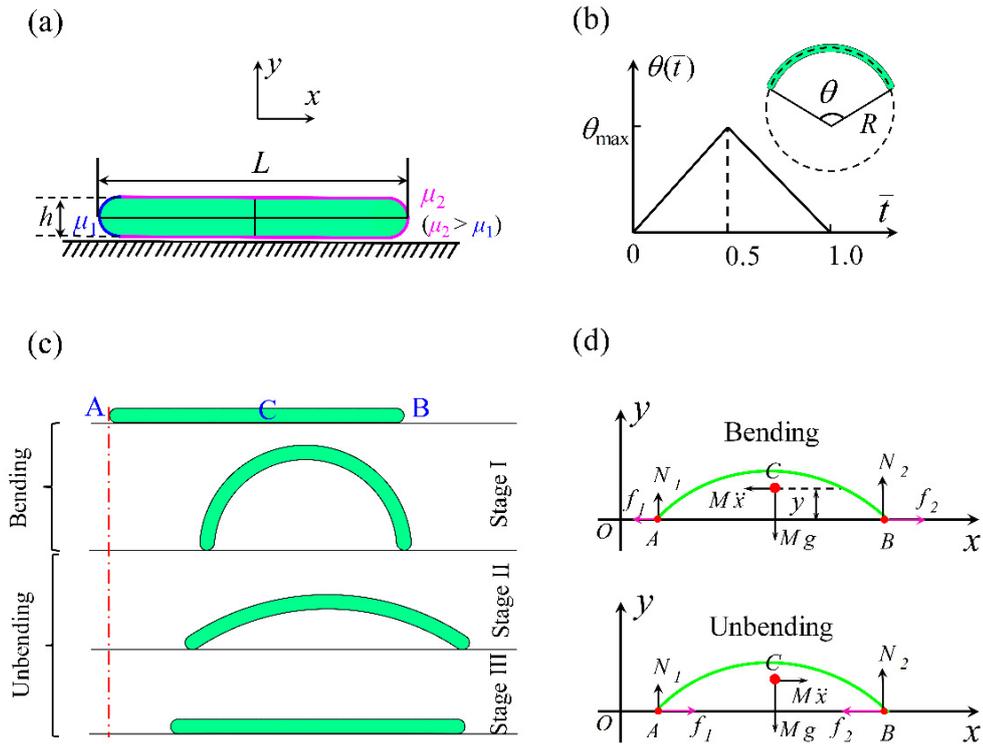

**FIG. 1.** (a) The worm-like crawling beam model with different frictional coefficients at the two ends of the contact surface. (b) The bending actuation of the beam is described by the central angle (inset) and the dimensionless time $\bar{t} = t/P$, where $P$ is the period. (c) Bending-crawling locomotion of the beam model in one cycle. (d) Force diagram of the beam (including inertia force) in the bending and unbending processes. Here the beam is represented by a curve and Point C is the beam's mass center. A video of this locomotion is available in Supplementary Video1.

The following kinematic analysis is based on the condition that the curvature varying rate is low. High varying rate may lead to "jump" of the beam model which will be discussed in the next section. Fig. 1c shows the bending-crawling locomotion of the beam in one cycle, which is divided into three stages. In the first stage the right end of the bilayer beam (point B) is stationary (i.e. "stuck") and the left end (point A) moves toward point B because of $\mu_1 < \mu_2$. With initial position of point A as the origin of the coordinate plane, the coordinate of the mass center (Point C) is $x_I = L - R\sin(\frac{\theta(\bar{t})}{2})$ $(0 < \bar{t} < 0.5)$ in the first stage. The second stage is the initial portion of the unbending process where the beam mass center decelerates; during this stage point A moves left and point B moves right. The kinematics of this process can be described by $M\ddot{x}_{II} = -\mu_2 N_2 + \mu_1 N_1$ $(0.5 < \bar{t} < \bar{t}_{II} < 1)$ (Fig. 1d). In the third stage, point B becomes stationary again and the mass center of the beam is at $x_{III} = x_{II}(\bar{t}_{II}) - [R\sin(\theta(\bar{t})/2) - R\sin(\theta(\bar{t}_{II})/2)]$ $(\bar{t}_{II} < \bar{t} < 1)$ where $x_{II}(\bar{t}_{II})$ is the coordinate of the mass center at the end of the second stage. The velocity of the beam becomes zero at the end of one cycle.

The crawling locomotion described above combines the advantages of simple structure, easy actuation and control (simple bending and unbending modes). However, its weakness is also notable. For example, the crawling efficiency is quite low that it takes several loading cycles to crawl one body length. Besides, the present mechanism relies on asymmetrical frictional forces, which requires some kind of asymmetry in material, geometry, or surface properties. The simple model presented in our earlier work [30] is also incapable of moving back, turning or climb over an obstacle, and modifications to the model are required for the crawling robot to achieve additional tasks, elucidated in Section 3.1.

## 2.2 In-plane snake-like slithering locomotion

As shown in Fig. 2a, a slender beam with length $L$ and square cross-section ($h \times h$) is placed on a rigid flat plane. The $z$-direction is opposite to that of the gravity. Here in the slithering model, all material, geometry, and surface properties are isotropic and uniform, including the friction between all surfaces of the beam and the ground. The beam is actuated by a periodic travelling sinusoid wave along the axis, and the time-dependent strain field applied to the beam's two lateral sides (Fig. 2a), are, respectively, $T(\bar{s},\bar{t}) = T_{max}\sin 2\pi(\bar{t}-\bar{s})$ and $T(\bar{s},\bar{t}) = -T_{max}\sin 2\pi(\bar{t}-\bar{s})$ [31]. Here, the dimensionless time is $\bar{t} = t/P$ where $P$ is the period, $\bar{s} = s/L$ where $s$ is arc length, and $T_{max}$ is the maximum strain applied. The bending curvature of the beam is derived as $\kappa(\bar{s},\bar{t}) = -\kappa_{max}\sin 2\pi(\bar{t}-\bar{s})$ (Fig. 2c) [31], where $\kappa_{max}$ is the maximum curvature located at the wave crest (red points in Fig. 2b).

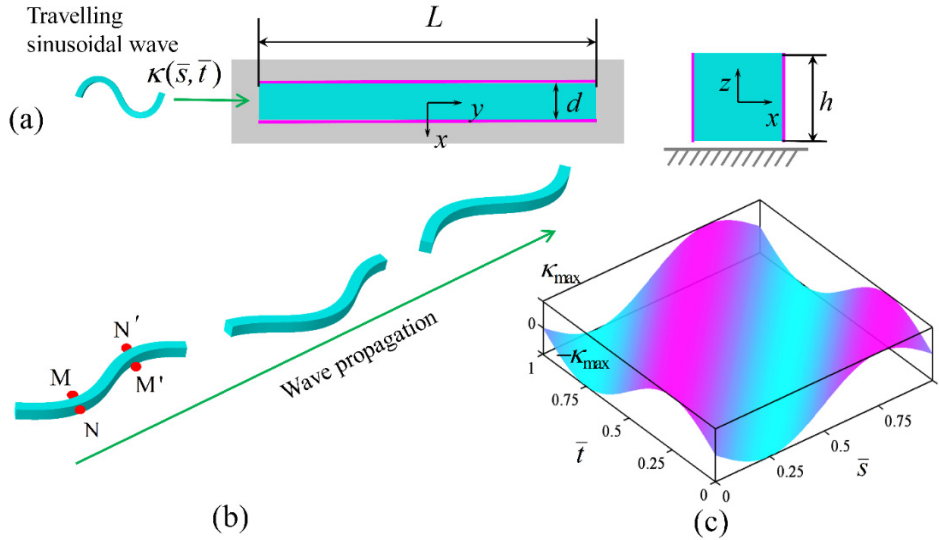

**FIG. 2.** (a) The snake-like slithering beam model on a flat rigid plane. (b) The snake-like slithering locomotion. (c) The travelling wave actuation of the beam.

Despite the uniform and isotropic properties, the snake-like robot undergoes locomotive motion toward a particular direction. Here, the symmetry is broken thanks to the differential contact areas at the maximum curvature regions due to the subtle Poisson's ratio effect. The Poisson's ratio dictates both the slithering speed and traveling direction. The contact area between the beam and the ground locates at the outer edge of the maximum curvature (point N and N′ in Fig. 2b) for negative Poisson's ratio, or at the inner edge of the maximum curvature (point M and M′ in Fig. 2b) for positive Poisson's ratio. During periodic wriggling of the beam, a rotational velocity field at the maximum curvature of the beam is generated. And because of the friction between the beam and the ground, the beam can slither in the same direction of the traveling wave for negative Poisson's ratio, or in the opposite direction of the traveling wave for positive Poisson's ratio. Besides, the slithering velocity for negative Poisson's ratio is much larger than that for positive Poisson's ratio, thanks to the larger moment (see Ref. [31] for details).

The steady-state slithering locomotion speed is

$$V_{\text{negative}} = \frac{L}{P}\left[\left(\frac{\kappa_0 L}{4\pi}\right)^2 + \kappa_0 \frac{h}{2}\right] \tag{1}$$

for negative Poisson's ratio, and

$$V_{\text{positive}} = -\frac{L}{P}\left[\kappa_0 \frac{h}{2} - \left(\frac{\kappa_0 L}{4\pi}\right)^2\right] \tag{2}$$

for positive Poisson's ratio.

The slithering locomotion mechanism described above is analogous to the moving wheels and has a relatively high moving efficiency (compared with the crawling mode). The moving direction can be changed simply by changing the wave propagation direction. Most importantly, it does not require friction anisotropy [9, 41, 42], nor lateral pushing against push-points [8]. The model can be escalated with more functions, such as turning or rolling, to be developed in Section 3.2.

## 3. Architectures of soft robotic locomotion

Based on the above two fundamental locomotion models, we present a series of locomotion modes achieving various locomotion functions, including crawling, slithering, rising, running, jumping, swimming, turning, turning over, helix rolling, wheeling, creeping, squirming, etc. All designs are enabled by simple mechanical principles without involving complex actuation or control algorithms. We focus only on the locomotion mechanisms and the detailed kinematic analysis is left for future studies.

**3.1 Locomotion architectures based on linear actuation**

New modes proposed in this subsection are based on the worm-like crawling model (Section 2.1), and demonstrated through FEM simulations using the commercial software ABAQUS. Appropriate mesh density is guaranteed through mesh convergence studies. The geometrical parameters for the beam model are length $L$ = 1.0 m, height $h$ = 0.05m, and width $d = h$ (for 3-D model). Friction coefficient between the beam and the ground is $\mu_1$ = 0.05 for the left end, $\mu_2$ =0.4 (unless otherwise specified). Self-contact of the beam surface has a friction coefficient of 0.4. The mechanical properties (e.g. Young's modulus and Poisson's ratio) of the beam can be in the range of that widely used in soft robots, yet their exact values are not essential for the locomotion mechanism in this subsection. More details can be found in our previous work [30].

**3.1.1 Rising and running**

As mentioned above, one major disadvantage for the worm-like crawling model is its slow velocity. If defining the movement efficiency ($\eta$) as the ratio between the moving distance per cycle to the maximum size of it [31], $\eta$ is about 0.2 for movement in Fig. 1. With only minor modification to the bending actuation of the beam, this model can achieve a relatively high locomotion efficiency. Different bending strains are applied to the left and right halves of the beam, i.e., the left and right halves of the beam have different $T_{\max}$ and thus different maximum excitation angles ($\theta_{L\text{-max}}$ and $\theta_{R\text{-max}}$, see Fig. 3b).

In the example in Fig. 3a-b, the left half beam's maximum excitation angle ($\theta_{L\text{-max}} = 7\pi/6$) is larger than that of the right half ($\theta_{R\text{-max}} = 5\pi/6$). Upon unbending, its right end rises up while the left end is contacting with the ground (Fig. 3b). In this case, the beam can move over half of its body length in one cycle ($\eta > 0.5$). Besides, since the head (right end) of the beam rises up in this process, such a locomotion form may easily crawl over an obstacle. Based on the model shown in Fig. 3b, if the curvature of the beam is released quickly after the beam reaches the maximum excitation curvature (for both the left and right half) (Fig. 3d), the model can jump forward with a very high speed ($\eta \approx 2$ in Fig. 3e), analogous to the running locomotion of the quadruped (e.g., the horse, tiger and cheetah). With the curvature of the beam rapidly decreasing, the ground generates a relatively large oblique upward force to the beam (Fig. 3c). This mechanism works both forward (Fig. 3d-e) and backward (Fig. 3f-g) despite of the different friction at the two ends of the beam. It is worth noting that this simple beam model does not have the excellent buffer capability as that of the quadruped. Thus when it lands on the ground, it may rebound several times until it becomes static. The impact between a soft robot and the ground was also observed and discussed for the combustion-powered "jumping robot" [43] developed by Bartlett et al. The ground impact behavior is not a focus in the present work thus will be discussed elsewhere. Future designs and fabrications involving running and jumping of the soft robot may need to elaborately consider the impact behavior.

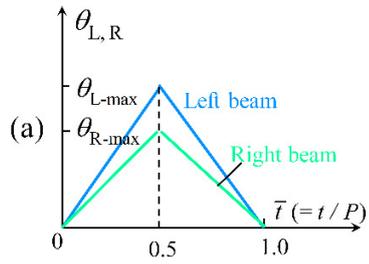

(a)

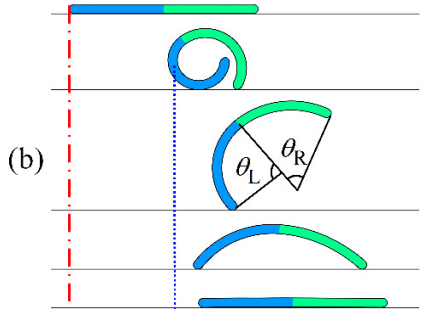

(b)

(c)

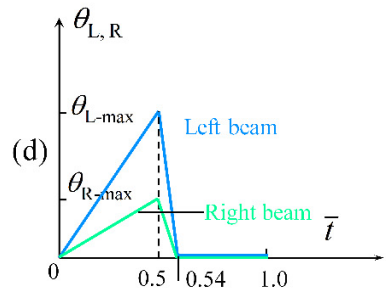

(d)

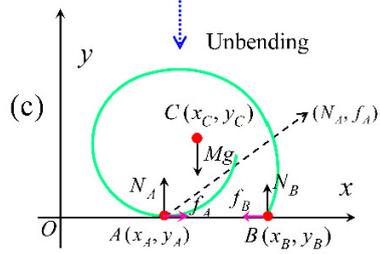

(f)

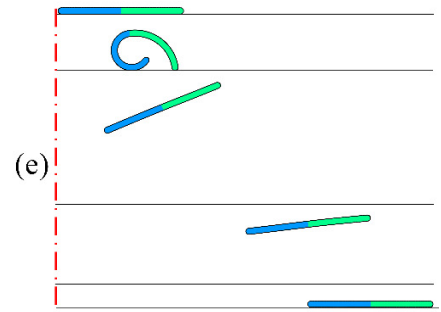

(e)

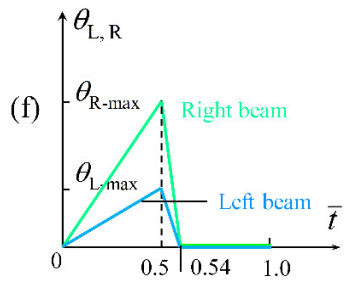

(g)

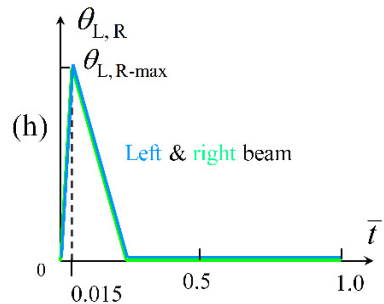

(h)

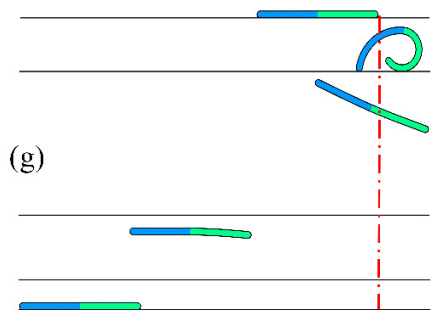

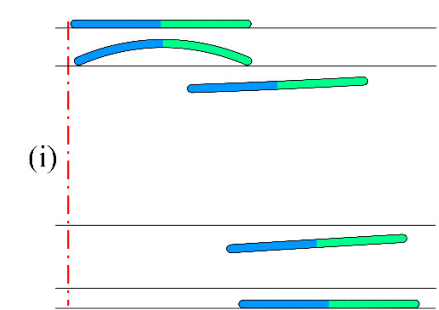

(i)

**FIG. 3.** The bending actuation and movement process for the rising-up locomotion (a, b), jumping forward (d, e) and backward (f, g) locomotion, and another jumping forward locomotion (h, i) in one cycle. The force diagram of the fully deformed beam (second panel of (b)) for rising-up locomotion is given in (c). $\theta$ is central angle of the bended left half or right half of the beam. $\theta_{L\text{-max}} = 7\pi/6$ and $\theta_{R\text{-max}} = 5\pi/6$ in (a-c), $\theta_{L\text{-max}} = 7\pi/6$ and $\theta_{R\text{-max}} = 3\pi/6$ in (d-g), and $\theta_{L\text{-max}} = \theta_{R\text{-max}} = \pi/6$ (h-i). The period $P=4$ s in (a-g), and $P=2$ s in (h-i). Videos of the rising-up (a, b), jumping forward (d, e), and another jumping forward locomotion (h, i) are available respectively in Supplementary Video2-4.

Fig. 3c shows the force diagram of the beam at the beginning of unbending during the rising-up locomotion. To lift the right end of the beam (Point $B$) off the ground ($N_B=0$ and thus $f_B=0$), the beam configure ($\theta_{L\text{-max}}$ and $\theta_{R\text{-max}}$), support and friction force at point $A$ ($f_A=\mu_2 * N_A$), and the release rate ($\lambda$, which affects $N_A$) should satisfy the condition that $N_A > Mg$ (gravity), and that the mass center (Point $C$) is located above the direction line (dashed arrow in Fig. 3c) of the counterforce (resultant force of $N_A$ and $f_A$) from the ground at point $A$. A large friction ($\mu_2$) between the beam and the ground helps to meet this condition because it shifts downward the counterforce direction. To further make the beam model "jump", a large curvature release rate ($\lambda$) is needed to induce large $N_A$. With Point $C$ much higher above the direction line of the counterforce at point $A$, e.g., as Fig. 3c shows, when the beam jumps into the air, it may flip over (several rounds) before it lands. In Fig. 3d-g, $\theta_{R\text{-max}}$ is chosen to be $\pi/2$ (instead of $5\pi/6$ as that in Fig. 3a-c) to lower the mass center and avoid flipping.

Another jumping mechanism is to bend quickly from the flat state (Fig. 3h, i), during which the contact forces from the ground are large in the vertical direction, and the differential friction coefficients at the two ends result in a net forward force. Thus a forward speed and an upward speed are generated before the model jumps off the ground. To obtain a large forward speed in this mechanism, a high friction coefficient at the beam's right end to the ground is needed ($\mu_2=1.0$ in Fig. 3i), like the "claw" of some animals.

**3.1.2 Turning over**

Similar to the bending configuration shown in Fig. 3a-c, if we further increase the maximum excitation central angle of the right half of the beam ($\theta_{R\text{-max}}$) to $\pi$ (Fig. 4a-b), the mass center (Point *C*) shifts leftward and is horizontally closer to Point *A* (Fig. 4c), thus when the beam unbends the contact point (Point *A*) moves right, and the beam may turn over to the left (Fig. 4b). Similarly, turning over rightward can be realized (Fig. 4d-e). Clearly, different locomotion modes (crawling, rising, jumping, turning over, etc.) depend on the combination of loading and surface parameters, such as the maximum excitation central angle ($\theta_{L\text{-max}}$, $\theta_{R\text{-max}}$), friction ($\mu_1$, $\mu_2$), and curvature release rate ($\lambda$), whose combination governs the positon of the mass center and the force diagram of the beam model. Detailed relationship between them requires full kinematic and dynamic studies similar to that in our previous work [30], which is left for future studies.

Similar to the turning-over locomotion, a caterpillar-inspired soft robot, called GoQBot [16], that mimics the ballistic rolling behavior of caterpillar, was developed for curling and rolling with a very high speed. This robot contains a hammer head with high-friction to the ground acting as a pivoting anchor, a composite body consisting of several mixtures of silicone rubbers, and two tail skids with low-friction that provide lateral stability and may also play a part in the curling and rolling locomotion. In other words, GoQBot has a non-uniform weight distribution along its body length, which is more complex than our current limbless model. Nevertheless, the increased movement efficiency with non-uniform weight distribution and highly rapid deformation may provide insights to improve our future model.

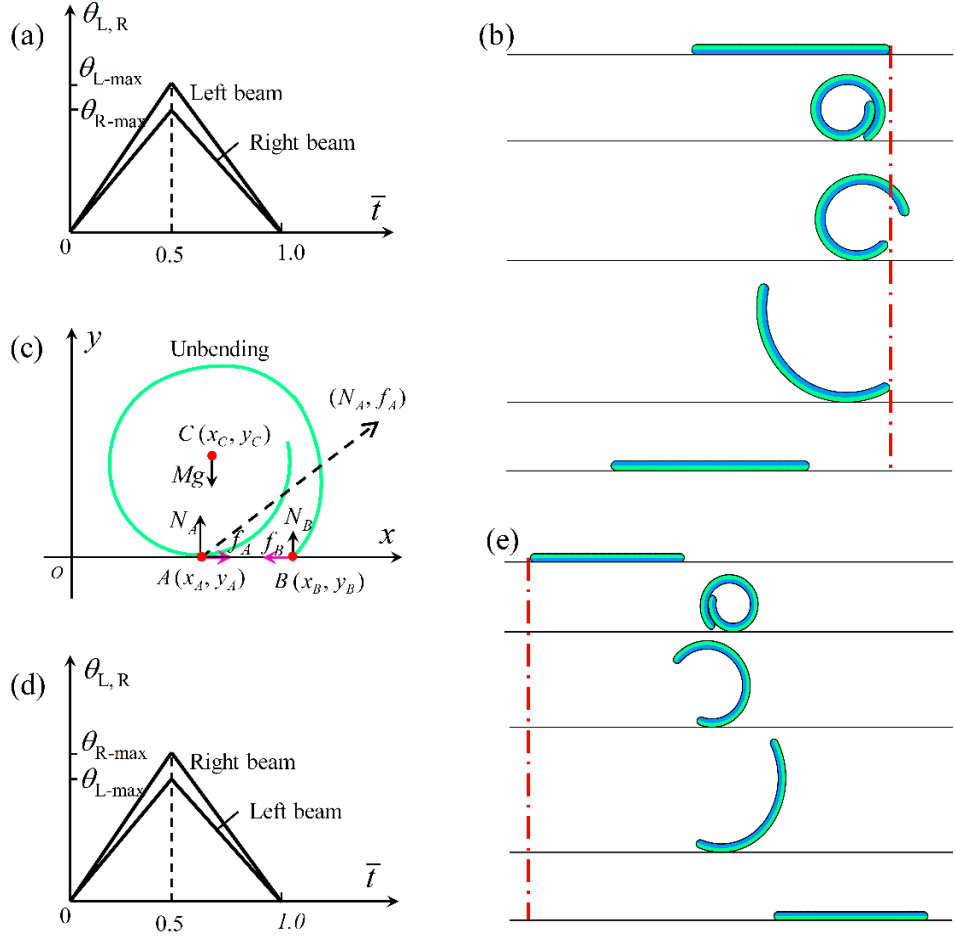

**FIG. 4.** The bending actuation and movement process for the turning-over leftward locomotion (a, b) and rightward locomotion (d, e). The force diagram of the fully deformed beam (second panel of (b)) for turning over leftward locomotion is given in (c). $\theta_{L\text{-max}} = 7\pi/6$ and $\theta_{R\text{-max}} = \pi$ in (a-c). $\theta_{L\text{-max}} = \pi$ and $\theta_{R\text{-max}} = 7\pi/6$ in (d-e). The period is $P=4.0$ s. A video of the turning-over rightward locomotion is available in Supplementary Video5.

### 3.1.3 Turning

Turning can be realized through multiple ways, such as a beam with in-plane curvature (Fig. 5a-b), or through two connected parallel beams with one moving faster than the other (Fig. 5c-f);

the latter can be realized through one long beam and one short beam subjected to the same loading parameter ($\theta_{max}$, Fig. 5c-d), or two same beams subjected to different loadings ($\theta_{Upper-max}$ and $\theta_{Lower-max}$, Fig. 5e-f). It takes dozens of cycles and large space for the beam with in-plane curvature to turn by 90 degrees (Fig. 5b). The parallel beam model has better time and space efficiency of turning (Fig. 5d, f). Besides, for the sled-like structure, through quick loading of one beam, it can roll over in the lateral direction (Fig. 5g-h).

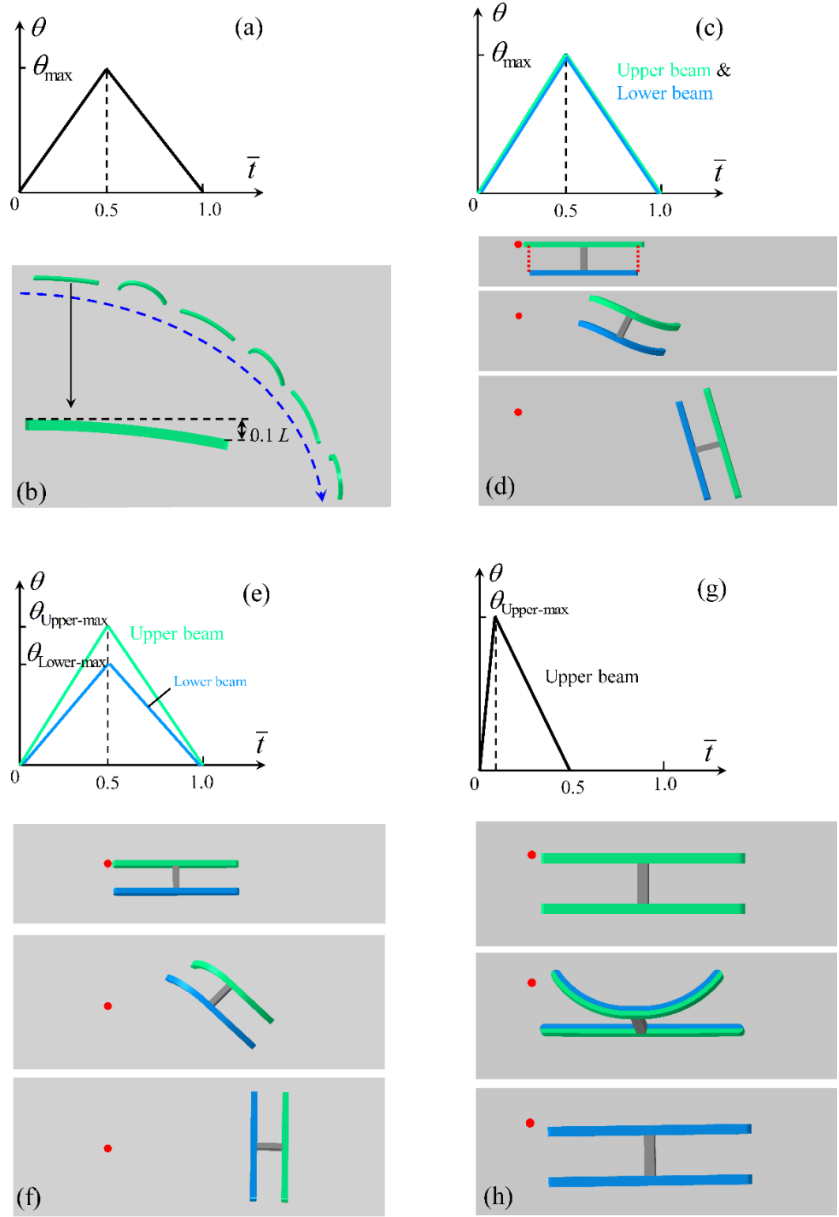
15

**FIG. 5.** The bending actuation and movement process for the turning of the in-plane bended beam model (a, b), turning of the sled-like model composed of one long beam and one short beam with the same loading (c, d), turning of the sled-like beam model composed of two same beams but with different loading (e, f), and lateral turning over of this model (g, h). The red point in each panel indicates the same spatial position. The maximum excitation angle is $\theta_{max} = 2\pi/3$ in (a-b). The maximum excitation angles for the upper and lower beams are $\theta_{Upper-max} = \theta_{Lower-max} = \pi$ in (c-d), $\theta_{Upper-max} = \pi$ and $\theta_{Lower-max} = 2\pi/3$ in (e-f), $\theta_{Upper-max} = 2\pi/3$ and $\theta_{Lower-max} = 0$ in (g-h). The period $P=2.0$ s. Videos of the above locomotion strategies are available respectively in Supplementary Video6-9.

## 3.2 Locomotion architectures based on travelling wave actuation

The following locomotion designs are based on the fundamental snake-like slithering model introduced in Section 2.2. The beam model here has a length of $L = 2.0$ m (unless otherwise specified), height $h = 0.05$ m, and width $d = h$. A negative Poisson's ratio of -0.3 is used since it is shown that the snake-like slithering model can move much faster with negative Poisson's ratio [31]. All friction characteristics are uniform with a friction coefficient of 0.3. More details can be found in our previous work [31].

### 3.2.1 Turning and spinning

In Fig. 6a, the beam model applies two complete time-dependent sinusoidal strain waves to the beam's two lateral sides, i.e., $T(\bar{s},\bar{t}) = T_{max}\sin 2\pi(\bar{t} - 2\bar{s})$ for one side and $T(\bar{s},\bar{t}) = -T_{max}\sin 2\pi(\bar{t} - 2\bar{s})$ for the other. Same as that in Fig. 2, the bending curvature of the beam center axis is $\kappa(\bar{s},\bar{t}) = -\kappa_{max}\sin 2\pi(\bar{t} - 2\bar{s})$, where $\kappa_{max}$ is located at the wave crest, and the beam slithers forward (Fig. 6b) at the same speed as that in Section 2.2. Typical movement efficiency ($\eta$) of such a fundamental locomotion mode is about 0.3 [31], depending on the maximum applied strain ($T_{max}$).

If the neutral plane of the beam is shifted (e.g. from black solid line to blue dash line in Fig. 6a) by adjusting the amplitudes of the applied strain field for the two lateral sides, for example, $T(\bar{s},\bar{t}) = T_{max}\sin 2\pi(\bar{t} - 2\bar{s})$ for one side and $T(\bar{s},\bar{t}) = -\alpha T_{max}\sin 2\pi(\bar{t} - 2\bar{s})$ ($0 \leq \alpha \leq 1$) for the other (Fig. 6a), then the snake model travels along a curved path (Fig. 6c). This turning mechanism is driven by differential excitation, and the mathematic description of the deformed shape, contact areas and velocity field, can be analyzed similar to our previous work [31].

To improve the turning efficiency (use less space to turn), in Fig. 6d-e, the strain field applied for the left half of the beam remains the same while the strain field for the right half is reversed. Thus the wave propagation direction for the left and right part of the beam is opposite (Fig. 6d). This leads to a spinning locomotion (Fig. 6e) with the center of the model static. Furthermore, increasing the amplitude difference of the two lateral surfaces (reducing the value of $\alpha$) or a wider beam can increase the differential driving moment, which may improve the space and time efficiency of turning.

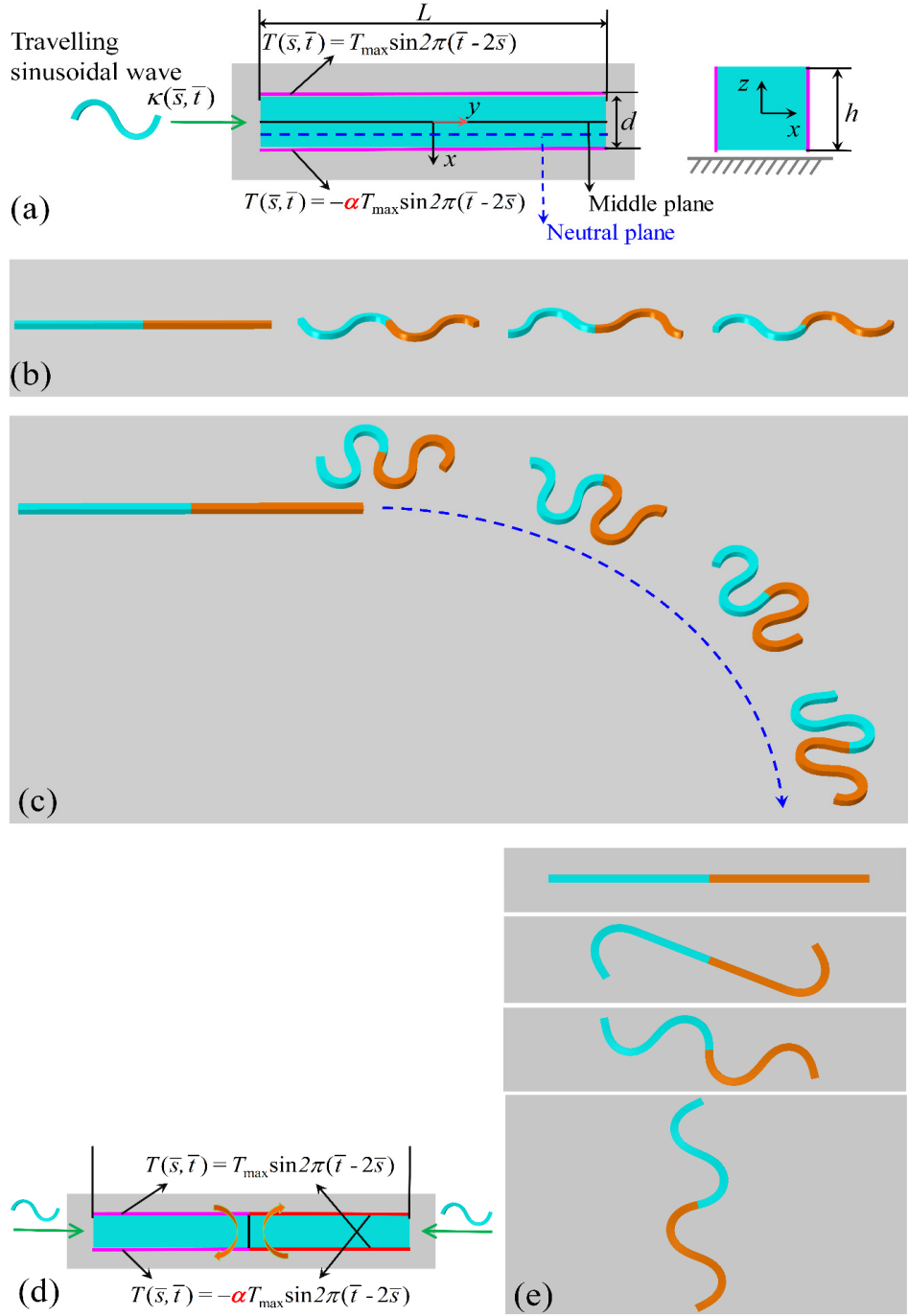

**FIG. 6.** (a) The travelling sinusoidal wave actuation for turning of the beam model. The amplitudes of the strain field for the two lateral surfaces are different, which leads the neutral plane of the beam deviating from its middle plane. (b) The two wave model moves forward in a straight line with $\alpha =1$, $T_{\max}=0.2$. (c) Turning of the beam with smaller strain amplitude for one lateral

surface ($\alpha=1/2$, $T_{max}=0.4$). (d) The opposite travelling sinusoidal wave actuation for the spinning locomotion of the beam model (e) with ($\alpha=1/3$, $T_{max}=0.3$). In all figures, the wave propagation direction is from the cyan part of the beam to its orange part. Videos of the above locomotion modes are available respectively in Supplementary Video10-12.

### 3.2.2 Wheeling

In the above sections about the in-plane snake-like slithering model, the strain field ($T(\bar{s},\bar{t}) = \pm T_{max}\sin 2\pi(\bar{t} - 2\bar{s})$) is applied to its two lateral sides. The system can be rotated by 90 degrees with the strain field being applied to the upper and lower sides, such that the deformation is now out-of-plane, as shown in Fig. 7a. The contact points between the beam and the ground are the crest of the deformed configuration (Fig. 7b). And the periodic movement of the waves has an effect of pushing the model forward analogous to the moving of wheels, which can be seen in Fig. 7b especially when focusing on the left end of the beam. Different from the previous work [31], the Poisson's ratio no longer affects the locomotion speed and direction in this manner. Furthermore, the initially undeformed beam can be a round ring placed vertically on the ground (Fig. 7c). With the periodic traveling wave propagates circumferentially along the beam, a self-actuated wheeling locomotion is obtained (Fig. 7c). Another advantage of this out-of-plane bending locomotion is that it can crawl up through two close parallel vertical walls (Fig. 7d), adding another movement dimension of this simple beam model.

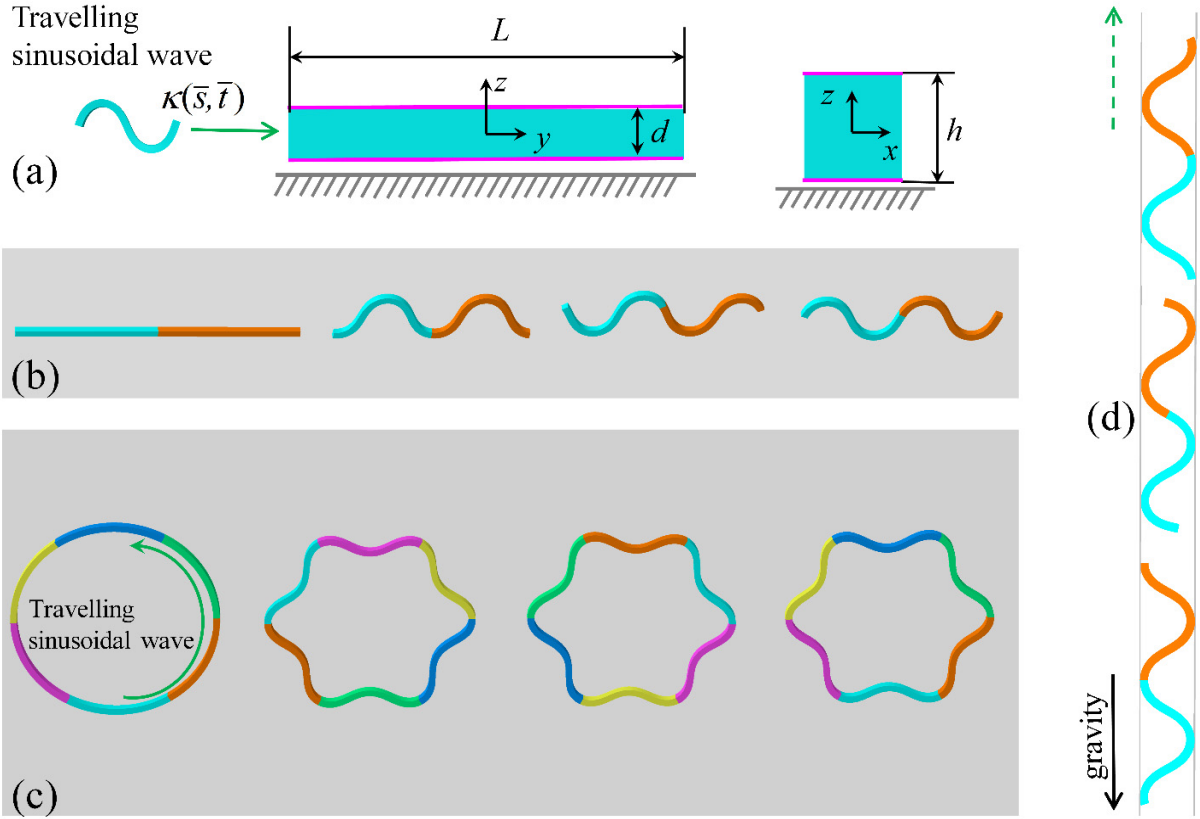

**FIG. 7.** (a) The travelling sinusoidal wave actuation for out-of-plane bending of the beam model. (b) The out-of-plane bending locomotion ($T_{max} = 0.2$) which is analogs to the movement of wheels. (c) The self-actuated wheeling of the initially undeformed ring model with out-of-pane sinusoidal wave travelling circumferentially ($T_{max} = 0.1$ for in inner surface of the ring, and $T_{max} = 0.2$ for the outer surface). (d) The out-of-plane bending locomotion can achieve a function of crawling up through a pair of parallel vertical walls ($T_{max} = 0.2$). Wave propagation direction is from cyan to orange. Videos of the above locomotion modes are available respectively in Supplementary Video 13-15.

### 3.2.3 Helix rolling

When the aforementioned two locomotion modes (the in-plane wiggling and the out-of-plane wheeling) are combined (Fig. 8a), the beam deforms helically and rolls (Fig. 8b). Note that the strain field applied to the lateral surfaces is identical to that mentioned in Fig. 6a and the strain field applied to the upper/lower surfaces has the same amplitude but with a phase difference of $P/4$ (Fig. 8a). During rolling, the beam moves laterally and forward with the lateral one faster. The proposed helix locomotion mode greatly expands this beam's adaptability to different terrains. For example, it can overcome gravity and crawl around a pillar or crawl through a pipe even when the pillar or pipe is vertical (Fig. 8c). The probability of such functions as climbing inside and outside of a pipe was previously mentioned by Wright et al. [44] with a "hard" snake model composed of several rigid modules connected by joints.

It is interesting that in order to climb upwards around a pillar, the excitation wave should propagate downward; whereas inside a pipe, the wave direction is same as the climbing one. A representative velocity field on one edge of the beam is plotted in Fig. 8d: when the beam is in a pipe, the beam's outmost area from its helix central line contacts with the pipe (indicated with red short lines in Fig. 8d). The velocities of these contact areas are generally opposite to that of the wave propagation, helping the beam to move in the same direction of the wave propagation. When around a pillar, the beam's closest area to its helix central line contacts with the pillar (blue circles in Fig. 8d), and the difference in contact area leads to distinct movements. Besides, the velocities of the red line areas are much larger than that of the blue circle areas, which explains the higher speed of climbing through pipe than that of climbing around pillar. More detailed mathematical descriptions of the velocity field of the beam (especially for contacting areas) for these locomotion patterns can be conducted following the framework in Ref. [31].

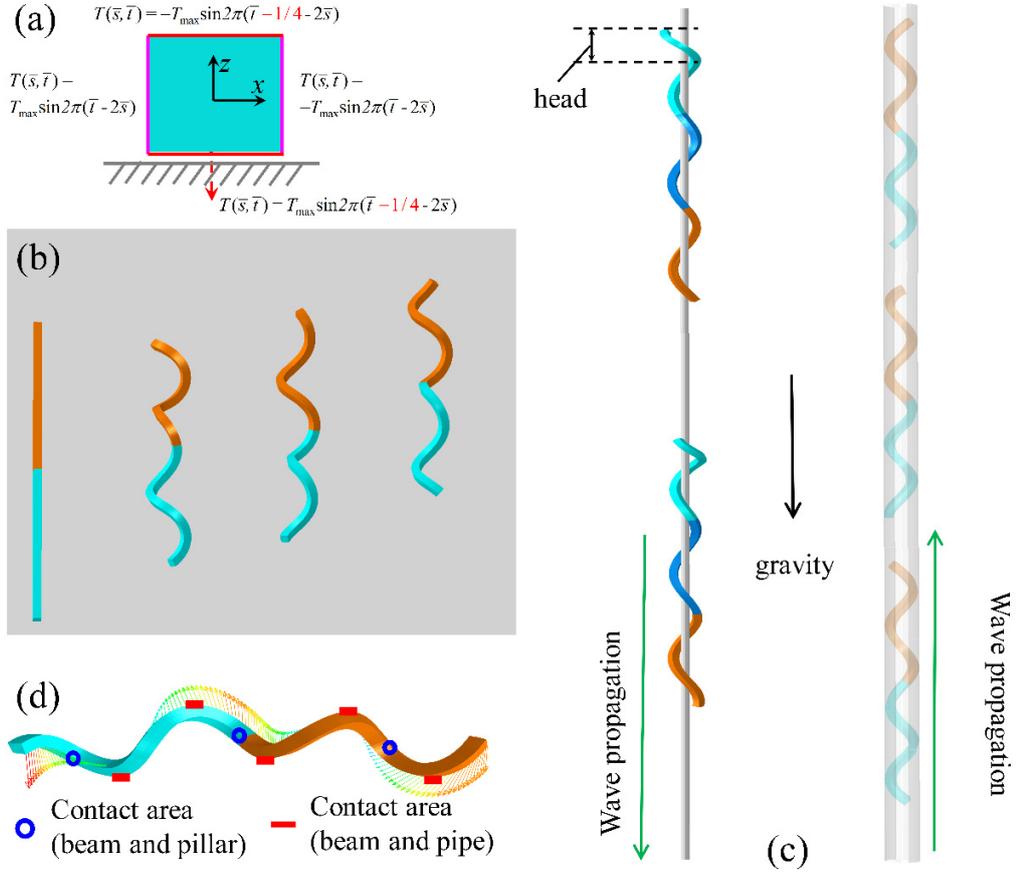

**FIG. 8.** The travelling sinusoidal wave actuation (a) for helix rolling locomotion of the beam model (b) with $T_{max}=0.2$. The time-dependent strain field is applied to the two lateral surfaces and the upper/lower surfaces with a phase difference of a quarter of the period $P/4$. (c) The helix locomotion enables the beam model to crawl around a pillar or through a pipe. As to the crawling around a pillar locomotion, for better twine between the beam and pillar, the beam contains 3 sine waves ($T_{max}=0.12$) and its head part (with length of half of the wavelength) is made to have larger deformation amplitude ($T_{max}=0.24$). Strain amplitude for the crawling through pipe model is uniform ($T_{max}=0.24$). (d) The FEM velocity field on one edge of the beam when it deforms in the vacuum environment. The blue circles indicate the contact area if this beam crawls around a pillar, and the red thick lines indicate the contact area if the beam is crawling through a pipe. Wave propagation direction is from cyan to orange. Videos of the above locomotion modes are available respectively in Supplementary Video16-18.

### 3.2.4 Turnover, 1-D squirming, and creeping

The excitation wave does not have to be sinusoidal. In this subsection, we illustrate the consequence of a square-wave strain field defined as $T(\bar{s},\bar{t}) = T_{max}H(\bar{t},\bar{s}) = T_{max}H(\delta,v)$, where $H(\bar{t},\bar{s})$ or $H(\delta,v)$ is a square-wave function with square-wave length $\delta$ and a traveling velocity of $v$. For example in Fig. 9a, $\delta = L/3$ and $v = \dfrac{\delta}{P/4}$, and outside the wave application area ($\delta = L/3$) no strain is applied. Thus, in each period of $nP \leq t < (n+1)P$ ($n \leq \bar{t} < n+1$) where $n$ is a positive integer, $H(\bar{t},\bar{s}) =1$ when $v\bullet(t-nP)-\delta \leq s < v\bullet(t-nP)$, i.e., $4(\bar{t}-n)-3\bar{s} \in (0,1]$, and for other values of $4(\bar{t}-n)-3\bar{s}$, $H(\bar{t},\bar{s})=0$ (illustrated in Fig. 9e). This square-wave field leads to a stable turnover locomotion (Fig. 9b), and its movement direction is opposite to the wave propagation direction.

In Fig. 9a-b, the strains on upper and lower surfaces have opposite signs. If the same sign is applied to both surfaces (Fig. 9c), the beam moves in a 1-D squirming manner akin to that of an earthworm (Fig. 9d). Note that in Fig. 9d, the Poisson's ratio is negative (-0.3) and the beam moves in the same direction of that of the traveling wave. For positive Poisson's ratio, the model moves in an opposite direction (data not shown).

Besides, the strain filed can be adjusted to contain both extension (in the middle) and contraction (on two sides) components (Fig. 9f-g), so as to obtain a creeping locomotion analog to the movement of some worms (Fig. 9h). Similar movements were achieved through a multi-gait soft robot [5] with undulation gaits, which is able to drive the soft robot underneath an obstacle.

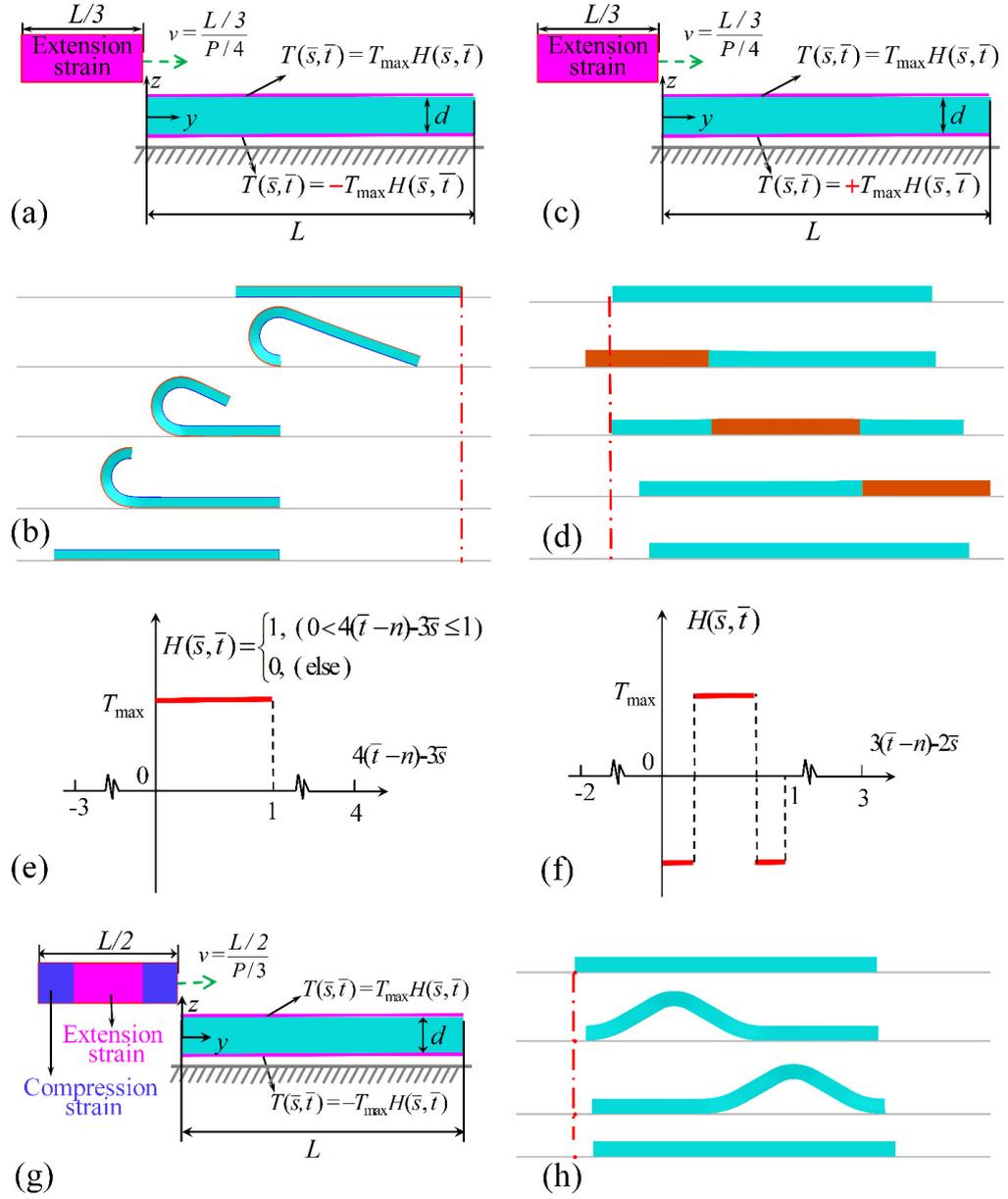

**FIG. 9.** (a) The beam model is subjected to the travelling square-wave strain field of $T(\bar{s},\bar{t}) = T_{max}H(\bar{t},\bar{s})$ for the upper surface and $T(\bar{s},\bar{t}) = -T_{max}H(\bar{t},\bar{s})$ for the lower surface ($T_{max} = 0.24$). For easy understanding, the traveling square-wave strain field is plotted as a rectangle above the beam model with length of $L/3$ and a velocity of $v = \dfrac{L/3}{P/4}$. (b) Turnover locomotion actuated by the strain filed described in (a). (c) The beam's upper and lower surfaces are subjected to the same strain field $T(\bar{s},\bar{t}) = T_{max}H(\bar{t},\bar{s})$ with $T_{max} = 0.24$, which leads to the squirming locomotion (d). The red part

indicates the traveling extension strain filed. Mathematic description and illustration of $T(\overline{s},\overline{t}) = T_{\max}H(\overline{t},\overline{s})$ used in (a-d) is shown in (e). (f-g) A new travelling square-wave strain field with both extension and contraction components ($T_{\max} = 0.16$), which allows a worm-like creeping locomotion (h). $L$=1.0 m in all figures here. Videos of the above locomotion modes are available respectively in Supplementary Video19-21.

### 3.2.5 Swimming and sand-swimming

A snake can swim on the surface of or inside water without the requirement of friction [45]. In Fig. 10a, the simulation is conducted using the smoothed particle hydrodynamics (SPH) method where the water is modeled as PC3D elements (3D continuum particle elements) in ABAQUS using continuum pseudo-particles. In this simulation, the beam model is floating on the water body with about half of its volume immersed in water. Similar results can be obtained with the beam immersed deeper. Its moving direction deviates a little from the initial axial direction (blue line in Fig. 10a), which may be caused by the initial perturbation related to morphological transition.

Besides, inspired by Maladen et al.'s study [46] of the sandfish lizard, *Scincus scincus*, we conduct another simulation to show that the present model is also capable to swim in sand (Fig. 10b). This simulation is conducted using the discrete element method (DEM) in ABAQUS where the sand particles are modeled as spherical particles with large radius (2 cm) to save computational cost. All contacts are assumed to have a friction coefficient of 0.3 and all normal force parameters (for particle-particle or particle-beam interactions) are uniform for simplicity. The beam is initially buried in sand at the depth of about its height (*h*). Similar results can be obtained with the beam buried deeper.

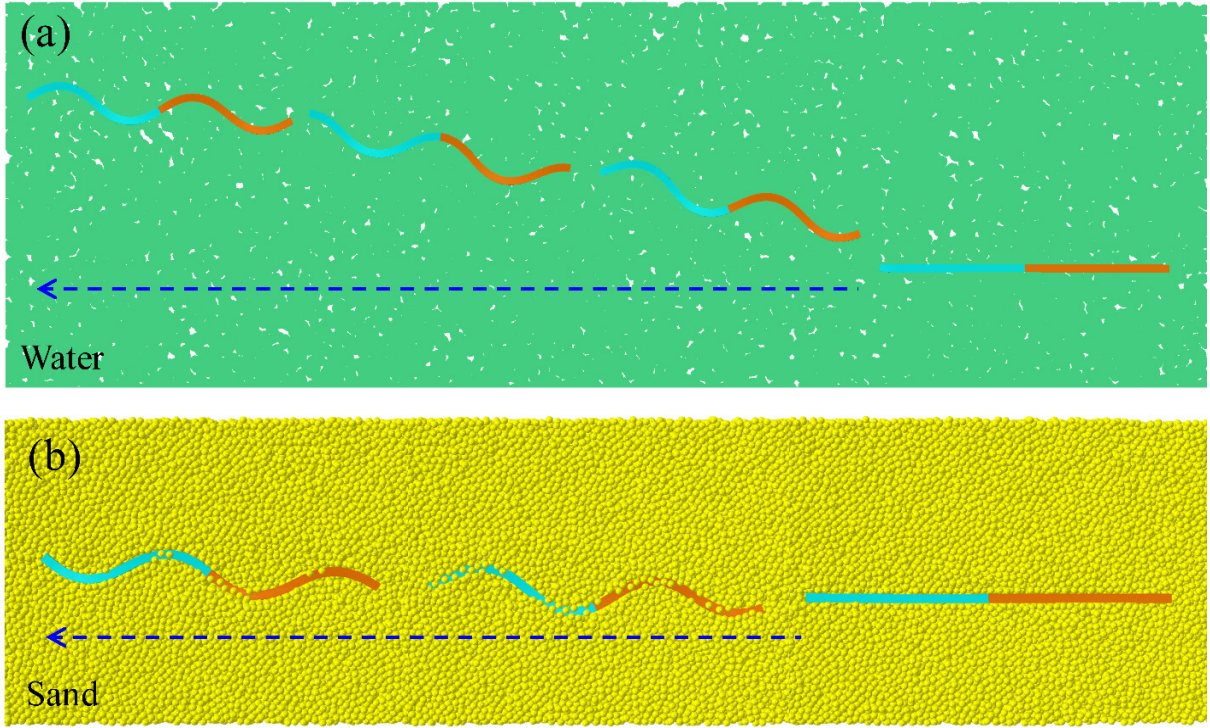

**FIG. 10.** Simulation results for water-swimming (a) and sand-swimming (b) locomotion. In water-swimming figure (a), only a layer of water body close to its upper surface is shown. In both figures, the propagation direction of the strain field is from cyan to orange with $T_{max} = 0.2$. To save computational cost, in the above two models, the water and sand "particles" have large radius (~2 cm). Thus the present simulations may only serve as proof-of-concept models. Videos of the above locomotion modes are available respectively in Supplementary Video22-23.

## 4. Comparison between present modes with other soft locomotion models

A comparison of the functional terrains of the present study and some previous biomimetic robots is shown in Fig. 11a. Ten representative terrains are listed in the ordinate, and the present beam model covers all of them, while the previous (soft) robots usually span 1-4 kinds. This justifies the wide adaptability of the present work.

Another important issue is the movement efficiency of the soft robot. Since actuation of soft robots is usually periodic, the movement efficiency is defined as the ratio between the moving distance per cycle to the initial body length [5, 9, 10, 12, 16, 46]. The deformation level is defined as the ratio between the initial size and the deformed size after actuation [12, 16]. It can be seen in in Fig. 11b that the present models excel in movement efficiency with less deformation energy, since most of them are able to deform quickly [16]. It is envisioned that the modes proposed herein may overcome moderately uneven terrains, jump (or rising / turnover) over an obstacle, roll around a wire or a tree, squirm through a small hole, etc.

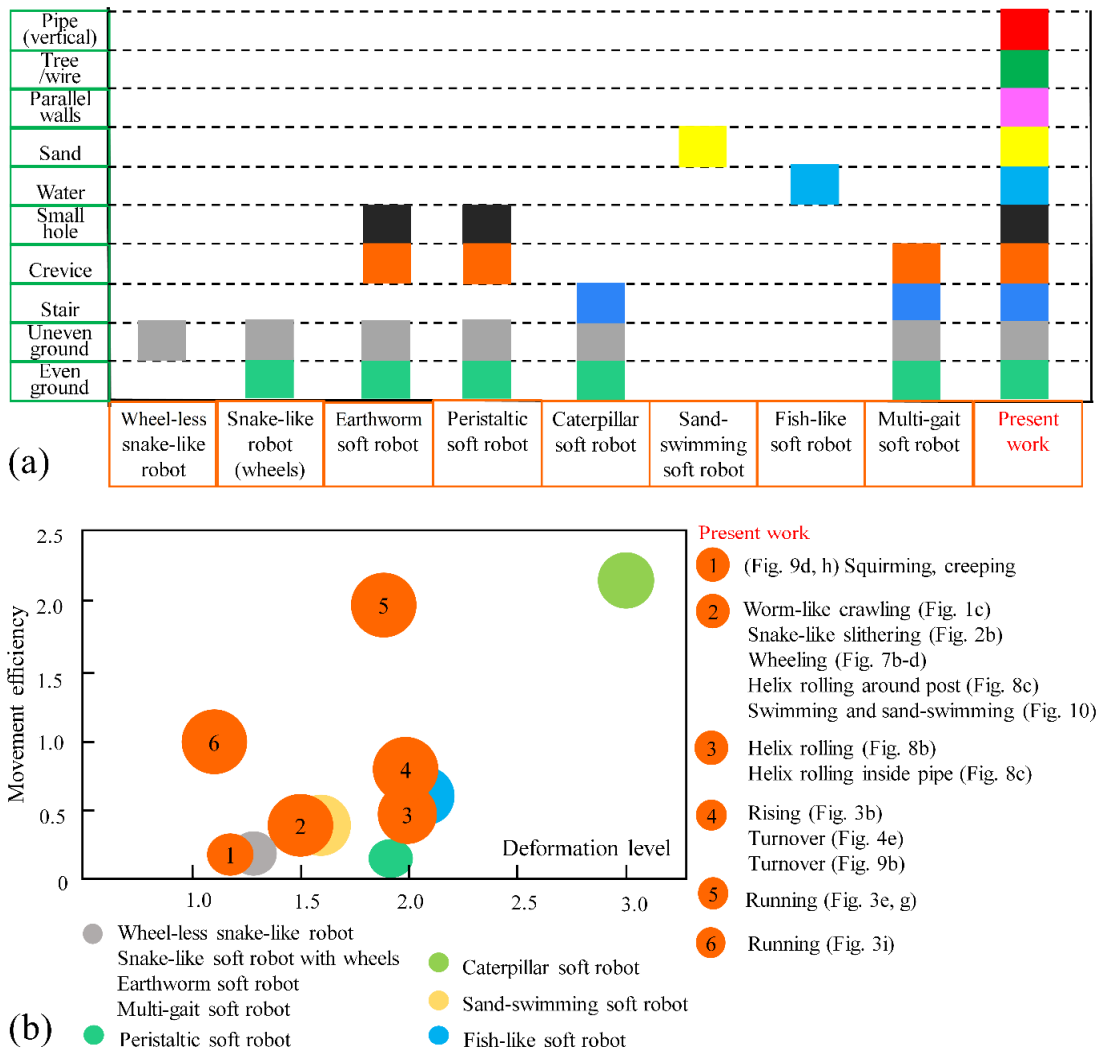

**FIG. 11.** (a) Comparison of the functional terrains of the present work and some representative (soft) robots. From left to right, these robots are the wheel-less snake-like (hard) robot [8], the snake-like soft robot with wheels [9], the earthworm soft robot [10], the Oligochaeta-inspired peristaltic soft

robot [23], the caterpillar soft robot [16], the sand-swimming soft robot [46], the fish-like soft robot [12], the multi-gait soft robot [5], and the present model. (b) Comparison of the movement efficiency and deformation level among the above mentioned biomimetic robots.

In Nature, snakes or earthworms have a high tolerance to partial body malfunction that when a part of its body loses excitation/ flexibility or broken, it can still move. Similarly, as illustrated in Fig. 12, the snake-like slithering model with partial malfunctioning body, as an example, does not lose much of its speed and flexibility during locomotion. Proper design of the control and power system for experimental prototypes may exemplify such tolerance.

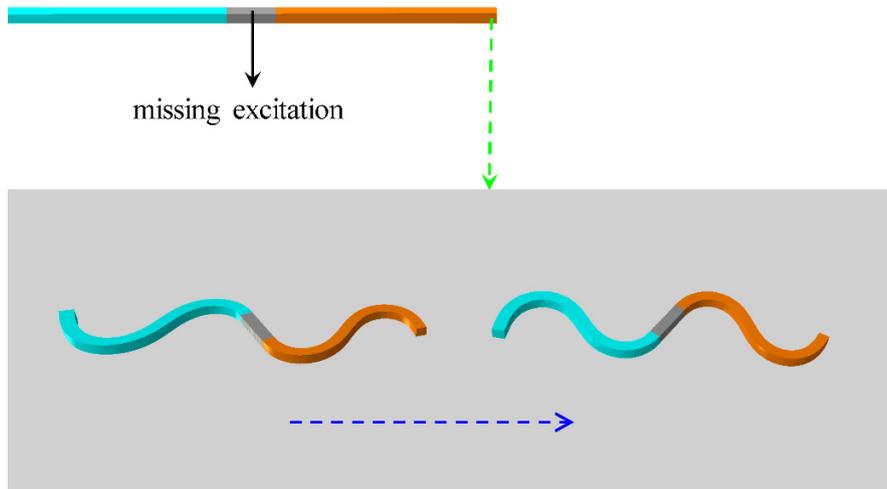

**FIG. 12.** A snake-like slithering model with partial body missing excitation, as an example, shows that the soft robot can be quite tolerable to partial body malfunction, analogous to that of an earthworm.

## 5. Practical implementation methods

The actuation methods and materials up to date [15] are sufficient for realizing the above locomotion strategies. Based on our previous work [31], we propose two representative strategies for soft robotic locomotion. The first one is a beam made of auxetic materials (structures) with discrete dielectric elastomer actuators (Fig. 13a) and the other has the pneumatic actuation chambers around the beam (Fig. 13b). Many other methods with different actuators (e.g. bimorph actuators [34-37]) may also work as long as the response is quick enough.

Schematically illustrated in Fig. 13a is the soft robot with dielectric elastomer actuators on all four lateral sides of the beam. Body of the robot is made of auxetic materials (structures) with negative Poisson's ratio. Note that materials with positive Poisson's ratio is also functional for all locomotion modes proposed in this work, only that for the snake-like slithering mode (Fig. 2 or Fig. 6b) its velocity is much smaller than that with auxetic materials [31]. Examples may include the simple cubic auxetic metamaterial [47] whose negative Poisson's ratio may reach -0.4 and may sustain a compressive strain up to 0.3, which suits well the present models. To actuate the beam, time varying voltages are applied to the dielectric elastomer actuators through its thickness, which expands/contracts in the axial direction to induce bending.

Fig. 13b shows the pneumatically actuated soft robot. The pneumatic chamber actuator in each cell can be controlled separately. With pressurized air applied, the chamber expands in both the axial direction (which generates the bending deformation) and the lateral direction. The effects of the structure of the pneumatic actuator (the number of chambers, wall thicknesses, etc.) on the mechanical deformation and locomotion is left for future experimental designs [31].

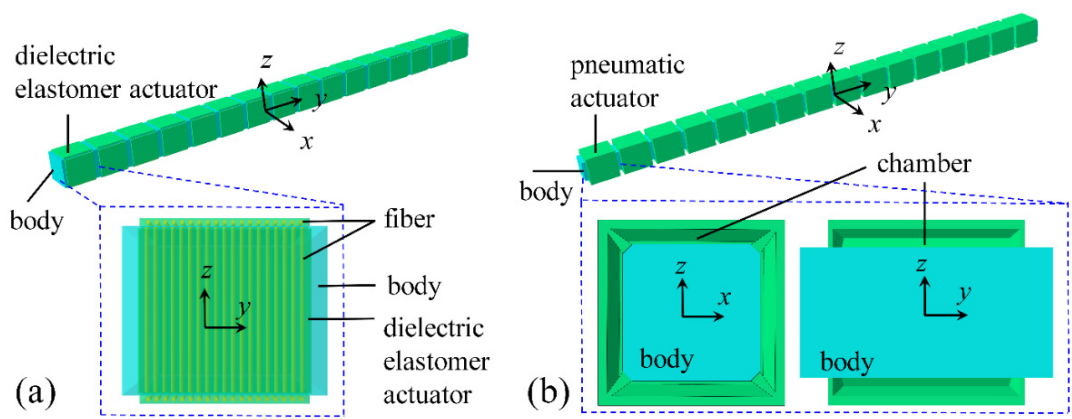

**Fig. 13** Schematic illustration of the soft robot with dielectric elastomer actuators (a) or pneumatic chamber actuators (b).

## 6. Concluding remarks

Series of locomotion strategies are developed based on simple mechanical principals. Underpinned by two fundamental locomotion mechanisms of beam model, i.e., the worm-like crawling and the snake-like slithering, over 20 new locomotion patterns are proposed, such as crawling, running, creeping, squirming, slithering, swimming, jumping, turning, spinning, turning over, helix rolling, wheeling, etc. With these locomotion strategies, this beam model is able to navigate a wide set of environments, having high adaptability and movement efficiency, and low deformation level. Examples include slithering or running on uneven terrain, swimming in water or sand, creeping or squirming through a small hole or crevice, crawling up around a tree, crawling through a (vertical) pipe or through two parallel close (vertical) walls. Note that even the malfunctioning of a part of the soft robot (e.g. missing excitation or loss of flexibility of part of the body) does not affect its overall capability. The present model is highly adaptive for severe, complex, and varying environments. These designs are all enabled by quite simple mechanisms echoing that certain nature's machineries may be designed simple. The study may further inspire series of sophisticated soft robotic locomotion studies and advance the development of soft robot prototypes, as well as the dynamics of soft materials.

**Acknowledgment**: X.C. acknowledges the support from the National Natural Science Foundation of China (11372241 and 11572238), ARPA-E (DE-AR0000396) and AFOSR (FA9550-12-1-0159). L. Z. acknowledges the support from the China Scholarship Council.